\documentclass[12pt,letter]{article}
\usepackage{graphicx}
\usepackage{latexsym} \usepackage{amssymb}
\usepackage{mathrsfs}
\usepackage{amsmath}\usepackage{amsfonts}
\usepackage{amssymb}
\usepackage{epsfig}
\usepackage{multirow}
\usepackage{layout}
\usepackage{here}
\begin{document}
\begin{center}
\begin{Large} 
\textbf{Horizontal symmetry in the algebraic approach of genetic code}
\end{Large}
\end{center}
\vskip.5cm
\begin{center} J.J. Godina-Nava\\
$^1$ Departamento de F\'isica CINVESTAV-IPN, Ap. Postal 14-740,\\
07000 M\'exico, D. F.; M\'exico\\
 jj@fis.cinvestav.mx
\end{center}
\vskip.5cm
\centerline{\bf Abstract}
\vskip.5cm
\begin{center}
\begin{minipage}{13cm}{\footnotesize
Using concepts of physics of elementary particles concerning the breaking of symmetry and grannd unified theory  we propose to study with the algebraic approximation the degeneracy finded in  the genetic code with the incorporation of a horizontal symmetry used in gauge theories to fit the contents of the multiplets of the genetic code. It is used the algebraic approch of  Hornos et. al. \cite{main,PRL71,PRE,MPLB}. We propose an example for the incorporation of horizontal symmetry to study mixtures of elements of the multiplets.   
} 
\end{minipage}
\end{center}
\begin{large}
\noindent

\Roman{section}
\section{Introduction}
\end{large}
 Because there are a countless number of similarities between disciplines, is that you get to set what is known as mutidiscipline. Area where certain techniques and tools of a certain discipline can be applied elsewhere in order to tackle problems of a complex nature.
This is the case of the biology and high energy physics (HEP). HEP combine a variety of concepts and technical tools commonly used in order to construct models to study matter and its components.
In biology there are problems that have long remained unanswered, especially dogmatic nature.
  However, proposals have emerged from HEP taking elements have been employed in biology in order to solve important problems that can helping to know more about a genetic disease like cancer, or a process such as DNA transcription or as the case that concerns us is the evolution of the genetic code that is very involved in the above mentioned problems.  Old problem still unsolved, is the search for a symmetry that responds to natural evolutionary behavior of the genetic code.
If we think of cancer as a disease caused by a translocation of nucleotides in a sequence.
The necessity to contemplate the genes by symmetry, leads us to think of a scenario where if it were possible to characterize the amino acids formed of 4 nucleotides chosen 3 at a time and their permutations, following a specific rule  unknown even in biology.  We could talk about this in the context of a certain unification of gene expression, motivating microscopic modeling more attached to reality.\\
In particle physics, we study an exciting scheme which had its origin in the dream of a physicist, Albert Einstein, who proposed a new formulation of all that was known in his time establishing his theory of general relativity. Thus, was born the search for unified field theories, theories with which one could describe with a single set of formulas the nature of matter. Grand unification theories are models that have been implemented in particle physics to unify the three known interactions of the standard model, each one of them characterized by a particular gauge symmetry.\\
The hypothesis is that these three interactions can be characterized by a single larger gauge symmetry and characterize it by only one coupling constant. The models that have been used to carry out this unification are semisimple Lie groups. The challenge is that by extending the symmetry, it is necessary to introduce more fields
and additional interactions and even extra dimensions of space, that from the experimental point of view have not been observed. The first successful model was the group SU(5), proposed by Howard Georgi and Sheldon Glashow \cite{georgi}, and preceded by the model of Paty-Salam
 \cite{pati}.  The essential part of the model is that the strong and weak interactions are described by the simple Lie groups SU(3) and SU(2) that allow to describe precisely the discrete electric charges. The weak hyphercharge is described by an Abelian symmetry U(1), which in principle allows the assignment of arbitrary charges, but there are however some restrictions on the choice of particle charges by theoretical consistency, in particular has to do with the cancellation of the so called triangular anomalies.\\
The quantization of charge allows to describe the particles by their content of electric charge they carry. This basic fact has been regarded as proof that the interactions have to be incorporated in a grand unified interaction described by a single group of higher symmetry containing the standard model. With this prescription automatically would be predicted the quantized nature and values ​​of the electric charges of all elementary particles.  The use of this type of model is used to constrain the theory based on experimental evidence. \\
In the biological size,  the structure and dynamics of DNA are the key to understanding the biological effects and the genesis of genetic diseases such as cancer. DNA is the molecule that encodes the information that an organism needs to live and reproduce. 
The genome of a cell contributes to the coherent modulation in time and quantity of proteins needed by living organisms in response to internal signals and the external environment. This behavior is the result of many factors acting at different hierarchical levels and time scales, not yet known in biology, including the regulation exerted by a family of proteins called transcription factors, which selectively bind DNA chain in order to activate or repress the expression of genes required in a coordinated manner  when those are needed. The gene expression is regulated at the level of mRNA and its role and abundance is regulated by these transcription factors. 
Modern techniques as the real-time polymerase chain reaction 
\cite{pfa} or the use of a green fluorescent protein gene as a reporter \cite{zas}, allows to obtain measurements at short sampling intervals, allowing to obtain time series with bigger and better temporal resolution for their study. \\
 These new techniques allow us to quantify the levels of expression with greater accuracy almost  without  noise measurements.  Despite these techniques, however, as well as the microarray, they share the drawback of only measuring the mRNA concentrations, instead of the corresponding proteins. The most common problem of these experimental techniques is to measure the present  level of protein concentration in the living cell.
These experimental improvements motivate the use of more sophisticated mathematical tools to address the problem. Thus the problem of understanding  the genetic code has been the subject of many models and still remains as a big challenge \cite{freeland, sella}. \\
Physical systems are governed by conservation laws. The law restricts the behavior of the system characterized by  symmetry. Symmetry as invariance to a specific group of transformations can apply the concept to problems in biology, such as genetic code to study diseases such as cancer.  The scenario here is of 64 codons, 4 nucleotide bases and 20 amino acids plus a stopping codon. The study of the physical properties of amino acids \cite{sitaramam} revealed a striking hierarchy. This suggested that for the breaking of the symmetry of the problem had to use certain  potential  to reduce the degeneration of the genetic code. 
The minimization of the effects of deterioration caused by accidental erasure or erroneous reading of a base-catalyzed sequence DNA polimererasa, showed a way to assign the stop signal to codons signiling stop biosinthesis \cite{jestin}.  Lehmann \cite{lehmann} found that the frequency of codon reading was closely linked to the reading system, being a symmetrical pattern of codons. The symmetries in the genetic code are of special interest because depending on these, you can get much information on the organization of code and therefore valuable information for gene expression \cite{PRL99}.  In the work of Hornos et. al. \cite{main,PRL71}, the genetic code was described using an irreducible  64-dimensional representation of a classical Lie algebra.\\
 The model reproduces the branching process with the correct multiplet structure of the genetic code and serve as a model to study its evolution. Is stated in \cite{PRL71,life} and summarized in  \cite{sachse} the conditions for the algebraic approximation to the genetic code from the previous experience in spectroscopy  \cite{PRL99},  where the states are associated with vectors of a irreducible representation from the generating algebra spectrum. The Hamiltonian is constructed as a linear combination of Casimir operators \cite{patera4} from generating algebra and its subalgebras.\\
But especially the conditions and terms under which is required as in grand unified models, which we remain with a residual symmetry: In the standard model is the $SU(2) \times U(1)$ we described above and  the hypothesis of Hornos et. al., is the called freezing or frozen accident, where they assume that the genetic code has evolved from forms much simpler with fewer amino acids and a degeneration much higher than that currently displayed. We can observe here an eloquent similarity between both theories and therefore we can perform parallel studies to find new information.\\
 Interesting efforts have been made since it was attempted to make a representation of the 64-dimensional space of codons  grouping the amino acids at the edges of a tetrahedral network 
\cite{tetra} to the new algebraic approximation in terms of simpletic $ Sp (6)$ algebra  
\cite{PRL71}, through the use of the Klein group structure \cite{montano},  model $SU(4)$ 
\cite{su4}, Galois theory \cite{galois} and many other approaches \cite{varios}.
 In what follows we will use the algebraic model $ sp(6) $ as the unifying symmetry
\cite{main, PRL71} and discuss the proposal of include a  horizontal symmetry   in the sense of a gauge  model of particle physics for study the mixing pattern,  particular relationship or rule of choice in the content of multiplets and their implication as fine tuning on the symmetry of unification. \\

\begin{large}
\section{Motivation}
\end{large}
We mention that the simplest model of Grand Unification was the SU(5). The smallest simple Lie group containing the  standard model was based on the model
\begin{equation}
SU(5)\supset SU(3)\times SU(2)\times U(1).
\end{equation}
The smallest irreducible representations of SU(5) are the $\bf{5}$ and $\bf{10}$. The assignment of content is that in the $\bf{5}$ are located the electric charge conjugate quark triplet type right down and leptons as left doublet of isospin. In the $ \bf{10}$ are located the six type quarks up, the left triplet colored quarks type down and the electron right.\\
This pattern is repeated for each of the three families of generations of matter known ($e$, $\mu $, and $\tau $), do not know why there are three families of leptons in nature and because precisely these. The amazing thing about this particular choice is that the model has no triangle anomalies. The Lie structure analysis shows that have few groups and representations that made to cancel the chiral anomaly. The content of particles considering the standard model is as follows 
\begin{eqnarray}
\bar{\bf{5}}=&(\bar{3},1) \oplus (1,\bar{2})\nonumber \\
\bf{\bar{10}}=&(3,2)\oplus) (\bar{3},1)\oplus (1,1) \\
\bar{\bf{24}}=&(8,1)\oplus) (1,3)\oplus (1,1) \oplus) (3,2)\oplus (\bar{3},2). \nonumber 
\end{eqnarray}
With this approximation is obtained a good value of Weinberg angle, the Higgs mechanism: $SU(5)\to SU(3)\otimes 
SU(2)\otimes U(1)$ is very economical, explains the problem of hierarchies and predicts proton decay with a half-life stable in the experimental range.\\
On the biological side, the amino acids are constructed as mentioned above using 4 nucleotides choosing 3 at a time and making all its permutations, generated a sequence of codons and the so called stop codon. The interaction between these is performed by two or three valence electrons. Each of these nucleotides in the DNA pairs with another antinucleotid, codons having a unique anti-codon. The rule of correspondence between the nucleotide triplets called codon in the DNA sequence and amino acids is known as the genetic code.\\
The DNA and RNA have the property that the sequence of triplets is accurate and the assignment of codons to amino acids form multiplets Figure \ref{gene}. What is not understood is why this happens in this way.  Following the work of Hornos et. al \cite{main,PRL71,PRE,IJMPB2003} where is set a systematic search of continuous symmetries in the genetic code. 
They have been discussed in detail breaking schemes based on the maximal subalgebras \cite{MPLB} of the symmetry symplectic algebra $\bf{sp(6)}$.
\begin{figure}
\begin{center}
\begin{tabular}{cccccccc}
\hline 
Amino acid& dim & & & &Codons& & \\
\hline
Arg & 6 & CGU & CGC & CGA & CGG & AGA & AGG \\
Leu & 6 & UUA & UUG & CUU & CUC & CUA & CUG \\
Ser & 6 & UCU & UCC & UCA & UCG & AGU & AGC \\
Ala & 4 & GCU & GCC & GCA & GCG &     &      \\
Thr & 4 & ACU & ACC & ACA & ACG &     &  \\
Val & 4 & GUU & GUC & GUA & GUG &     &  \\
Gly & 4 & GGU & GGC & GGA & GGG &     &   \\
Pro & 4 & CCU & CCC & CCA & CCG &     &   \\
Stop& 3 & UAA & UAG & UGA &     &     &   \\
Lle & 3 & AUU & AUC & AUA &     &     &   \\
Lys & 2 & AAA & AAG &     &     &     &  \\
Cys & 2 & UGU & UGC &     &     &     &   \\
His & 2 & CAU & CAC &     &     &     &      \\
Asp & 2 & GAU & GAC &     &     &     &    \\
Glu & 2 & GAA & GAG &     &     &     &    \\
Tyr & 2 & UAU & UAC &     &     &     &   \\
Phe & 2 & UUU & UUC &     &     &     &   \\
Asn & 2 & AAU & AAC &     &     &     &   \\
Gln & 2 & CAA & CAG &     &     &     &  \\
Trp & 1 & UGG &     &     &     &     &  \\
Met & 1 & AUG &     &     &     &     &  \\
\hline
\hline
\end{tabular}
\end{center}
\caption{The codon representation for the fundamental amino acid multiplets}
\label{gene}
\end{figure}
We must be careful do  not to alter the circumstances concerning the remnant symmetry that must be to safeguard the survival of the genetic code. 
An important detail in the Hornos work, is the way of how to assign the category of fermion or boson in the multiplets content of. In particle physics, there is a rule well established on the basis that its nature is determined by the quantum numbers describing their actual physical condition and thus how these particles behave in nature under certain gauge interactions. \\
In the case of codons in the genetic code, this assignment is difficult, because the characteristic of self-organized complex dynamic system of the same makes it very difficult to perform. What is done is actually implement the tools and use experimental data at hand 
\cite{varios}. In our case where the purpose is to find a symmetry for the genes in search of establishing patterns of expression to study diseases.  The  Hornos's contribution has provided the opportunity to be involved effectively with a new tool to treat a biological problem very important that has not solved for decades. The use of an algebraic approximation to the problem provides a valuable tool to study it.\\
In this sense, decoding fermionic or bosonic nature of the elementary components of the genetic code is important, therefore models have been proposed as the fermionic algebras $sp(6)\supset U(3)$ of Hecht \cite{hetch}, to study the fermionic foundation in the bosonic interaction model built by Arima and Iachello \cite{lachello} as a dynamic symmetry model. 
In particular, we propose to implement as in elementary particle physics, a symmetry that allows us to study the pattern of mixing between the components of the multiplets. This is to implement an extra horizontal symmetry to the symmetry of the genetic code as has been done by Hornos et. al. \cite{main,PRL71} in a grand unified theory for genes and follow the references \cite{horizontal,horizontal1} to implement it.\\
Hornos et. al. \cite{PRL71}, employed quaternion matrices and reproduces the genetic code breaking  in four stage the symmetry: $sp(6)\supset sp(4)\otimes su(2)\supset su(2)\supset su(2)\otimes su(2)\otimes su(2)$, where second and last $su(2)$ break go to $o(2)$ breaking the degeneracy of the code.\\
However, the symmetry $sp(6)$, has been considered as a continuous symmetry too large to describe the genetic code. Also the decomposition of the symmetry sp(6) does not reflect the actual process of refinement temporary codon recognition \cite{wilis}. Hence some discrete symmetries are  used \cite{finley}. \\
But the question now is that there is a conflict in the way in which the assignments are made of fermions and bosons. In reference  \cite{kent}, is established that the symplectic group sp(6) is part of a greater unification with the group SO(14) which also includes the group G(2), however the subset sp(6) contains greatest theoretical wealth. In any case, the idea is to have the 6-dimensional representation or three 2-dimensional in order to adjust the 20 amino acids.
We can see how delicate is the assignment of boson or fermion  to the content of multiplets, because there is a theorem in particle physics \cite{raifeartaigh} that rules out  the possibility of combining fermions and bosons in an multiplet  or in a irreducible representation of a Lie group. This can only be done using super Lie groups as it has been doing Forger et al \cite{forger-susy} and Bashford \cite{bashfords,bashfords1} introducing both commutative and noncommutative variables. For that as set forth in  \cite{kent},  in the baryon octet model SU(3) the adjoint representation 21 contains isospin multiplets with both integer and half integers, but all the baryons are fermions. Same is true in the meson nonete where all are bosons. This reinforces our proposal to include a horizontal symmetry to set this rule and allow freezing as also the symmetry of custody. The proposal is to continue the work of several authors in this regard
\cite{varios,mogardo} and assisted by the biological part \cite{ignacy,rietman,miloje,vitaly,andrew} establish the rule more plausible to use.\\

\section{The model sp(6)}
The algebra $\bf{sp(6)}$ \cite{hetch} has an 64 dimensional irreducible representation with higher weights $(1,1,0)$ according to the Cartan classification theorem using the generated base from the fundamental weights. In terms of the  diagonal base ${e_1, e_2, e_3}$ that describes the coordinates of roots and weights,
\begin{eqnarray}& &
(2,1,0)~~~~~~\mbox{highest weight},\nonumber \\& &
(2,0,0),(0,2,0),(0,0,2)~~~\mbox{long roots},\\& &
(1,\pm 1,0),(1,0,\pm1),(0,1,\pm 1)~~~~\mbox{short roots}.\nonumber
\end{eqnarray}
Calculating the scalar products with the highest weight (2,1,0) gives the length $m(\alpha)$ and subtracting are obtained the weights. Through the application of the transformation of Weyl group is generated the complete diagram of weights for the representation of sp(6) formed by 24 weights
\[
(\pm 2,\pm 1,0),(\pm 2,0,\pm 1),(\pm 1,\pm 2,0),(\pm 1,0,\pm 2),(0,\pm 2,\pm 1),(0,\pm 1,\pm 2),
\]
of lenght $\sqrt{5/2}$, there are also 8 weights type $(\pm 1,\pm 1,\pm 1)$ of lenght $\sqrt{3/2}$ and finally 6 weights $(\pm 1,0,0),(0,\pm 1,0),(0,0,\pm 1)$ of lenght $\sqrt{1/2}$. All these weights constitute a regular polihedro in a 3-dimensional space where it was possible to adjust the genetic degeneration of the code by Hornos et al \cite{main,PRL71}. The branching for decomposing the codons representation employed a canonical chain
\begin{equation}
sp(6) \supset sp(4)\otimes su(2)\supset su(2)\otimes su(2)\otimes su(2),
\end{equation}
here a very important part for the model of Professor Hornos is that depending on the assignment of their respective choice of algebra $su(2)$, likewise is assigned the fermionic or bosonic  nature of. This means \cite{main,PRL71} that for a spinor representation of a specific SU(2), will be designated as a bosonic representation if it has spin 1 under the group 
SU(2), escalar if it has spin 0 and fermionic if it has semi-integer spin.  In fact in \cite{main,PRL71}, is established that half of amino acids will be considered fermions and bosons will be the other half.\\
 The second part of the breaking \cite{main,PRL71} is made according the chain
\begin{equation}
su(2)\otimes su(2)\otimes su(2)\supset su(2)\oplus o(2)\oplus su(2)\supset su(2)\oplus o(2)\oplus so(2),
\end{equation}
The quantum states are governed by the highest weights $(k_1,k_2)$ under the 
$sp(4)$ and the highest weights of the spin $(2s_1,2s_2,2s_3)$ of any of the $su(2)$, are defined with this and the values of 
$(2m_1,2m_2,2m_3)$ of the magnetic quantum number determining the states
\begin{equation}
\mid k_1,k_2;s_1,s_2,s_3;m_1,m_2,m_3>.
\end{equation}
The methodology used until here \cite{MPLB} argues that in the analysis of finite groups is necessary to define a partial symmetry breaking that allow to remain with a residual degeneracy in the final stage to keep the contents of multiplets. Here is plausible to introduce the idea of a horizontal symmetry to keep the accidental degeneracy by incorporating it over the isotopic components because what is required is that this will be a maximal subgroup. Another interesting detail is given to looking for the study of mutations \cite{PRL71} in the second state we find 3 breaking 
$SU(2)$ that can help in the task. The proposal for a horizontal symmetry in the sense explained in the following section is that there is great ambiguity in the assignment of the doublets, quadruplets, and sextuplets therefore if one could incorporate a rule that relates them without alter the multiplets would be helpful. The discussion in \cite{scielo} about that the evolution of codons represented by a dynamical system can be considered as unique, it gives confidence in the evolutionary dynamics of amino acids. It is reported there that there is no mixing of different amino acid found in the different attractors. Therefore is recognized the complicated nature of the 6 degeneration of the amino acids.

\section{Horizontal approximation}
\begin{figure}
\centering
\resizebox{1.0\columnwidth}{!}
{\includegraphics{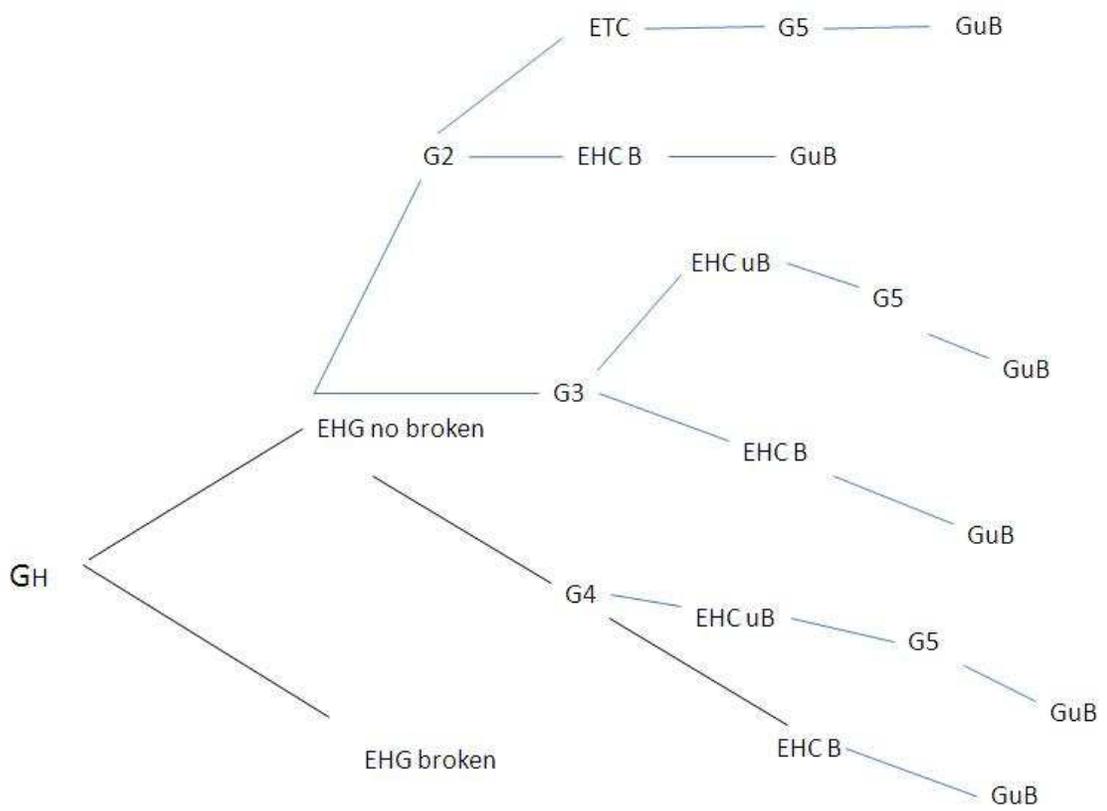}}
\caption{SSB for the horizontal symmetry}
\label{fig}
\end{figure}

As we have seen the concept of symmetries has been of great importance in Physics and given the parallelism of the problems we are attacking, is regarded plausible once again employ a concept of the theory of elementary particle physics to find a guide on the biological problem that concerns us, namely, considerations of symmetry we have helped to create the standard model and when asked the question of whether the same would achieve to unravel the problem of three generations of fermions known, emerged the idea of horizontal symmetry. That is, a way to fix the content of multiplets in order to have a more realistic scenario from an experimental point of view. In the following discussion will use the dissertation of reference \cite{horizontal1} to establish the technology to use.
Ordinarily in particle physics are matrices that characterize the properties of the particles under study. They are generally mass matrices for the respective sectors studied, ie quarks up, down, charged fermions, neutrinos. These matrices are diagonalized using unitary matrices. The idea is to build the horizontal symmetry to commute with existing matrices.  This will give you the same eigenfunctions, the same eigenvectors employees from all fields. In the first instance should have a horizontal symmetry for each quark sector we find.
This symmetry can be present at high energies and to break at current energies allowing precisely this residual symmetries or relationships between components of the multiplets. In this case it is clear that we require the use of a finite group to describe this symmetry, so it requires \cite{horizontal1}
${\cal G}=\{F,G \}$ with $F$ and $G$ as the residual symmetries in both sectors of quarks and the work would be to find the relation between the mixing matrix $U$ and the horizontal group ${\cal G}$. But the $U$ must be given by experiments, then the horizontal group ${\cal G}$ will have to find from the remnant symmetries $F$ and $G$. Thus, it must be found which are the possible ${\cal G}$'s for a given $U$ and which are the potential $U$'s for a given ${\cal G}$, how construct the invariant terms  under the ${\cal G}$ and how spontaneously break them in order to obtain the proper mixing pattern. In the Biological case as we have a basis where we can diagonalize simultaneously.  The eigenvalues of $F$ and $G$ can be chosen degenerate or not.
This reminds us of the palindromes of the genetic code and that \cite{tetra,montano,su4,varios,robersy1,tidjani,castro,robersy} some researchers have tried to restrict at peculiar symmetries. In our case it is important find them because we require to find their typical pattern. Let's look at how this could implement \cite{barbara}.
Symmetry groups allowed to have many links to adjust and higgses expected values ​​to adjust the model under study. With fine tuning of the parameters we can reach attractive results. It is expected that the horizontal symmetry is broken spontaneously and as you add a group of grand unification, this will establish the general details with the characteristics of both masses and mixing between components of the multiplets, the incorporation of horizontal symmetry will be made only make fine adjustments to the theory, as it induces mixing and mass differences really calculated. The additional Higgses introduced to break the horizontal symmetry, will contribute in some way and will have to be dimensioned. Is adjusting these parameters allowing us to use experimental data to implement models with different symmetries. In fact, using a horizontal symmetry instead of degeneration imposed between components of a multiplet, is a very good motivation to use \cite{nir}. Consider an example.
Suppose we have a symmetry group $G=SU(N)_I\otimes SU(N)_{II}$, restricted to $N=3n_f+ N^{'}_c N^{'}_F$ quark flavors and colors. In the representation of the symmetry,
\begin{equation}
\Delta \to (N, \bar{N}),~~\Delta=\sum_{n=0}^{N^2-1}\delta_n\delta_n,~~\lambda_0=\sqrt{\frac{2}{N}} \bf{I}_{N\times N},
\end{equation}
\begin{equation}
SU(N)_I\otimes SU(N)_{II}\to SU(3n_f)\otimes SU(N^{'}_cN^{'}_F)_{I+II}\otimes U(1)_{{\small \Sigma}}^{I+II},
\end{equation}
where the break is performed with 
$<\Delta>=<\delta_0>\lambda_0+<\delta_{\Sigma}> \lambda_{\Sigma}$ \\
con  $\nu=\sqrt{\frac{N^{'}_cN^{'}_F}{3n_f}}$

\begin{eqnarray*}
\lambda_{\Sigma}=\sqrt{\frac{2}{N}}   
\left[ \begin{array}{cc}
\nu & \cdot \\
\cdot & -\nu^{-1} \\
\end{array} \right]. \qquad
\end{eqnarray*}
The generators $T_a=\frac{1}{2}\lambda_a$, $\lambda_a\lambda_0=\lambda_0\lambda_q=\sqrt{\frac{1}{N}}\lambda_a$ for all a; the trace $Tr \lambda_a \lambda_b=2\delta_{ab}$, the $\lambda_f$ are the generators of $SU(3n_f)$, the $\lambda_F$ are the generators of $SU(N^{'}_cN^{'}_F)$ fulfilling the relationship
\begin{eqnarray}
\lambda_f\lambda_{\Sigma}=\lambda_{\Sigma}\lambda_f=\nu \sqrt{\frac{2}{N}}\lambda_f\\\lambda_F\lambda_{\Sigma}=\lambda_{\Sigma}\lambda_F=-\nu^{-1} \sqrt{\frac{2}{N}}\lambda_F.
\end{eqnarray}
For the interaction
\begin{equation}
D_{\mu} \Delta = \partial_{\mu} \Delta -ig_I Y^I_{\mu a} T_a \Delta+ig_{II} Y^{II}_{\mu a} \Delta T_a.
\end{equation}
The expectation value is
\begin{equation}
<\Delta> = \delta_0\lambda_0+\delta_{\Sigma}\lambda_{\Sigma}.
\end{equation}
Disaggregating
\begin{eqnarray}
D_{\mu} <\Delta> &= -i \frac{g_I}{2} Y^I_a \lambda_a (\delta_0\lambda_0+\delta_{\Sigma}\lambda_{\Sigma})+ i \frac{g_{II}}{2} Y^{II}_a  (\delta_0\lambda_0+\delta_{\Sigma}\lambda_{\Sigma})\lambda_a \nonumber \\  
&=-i\frac{\delta_0}{2}(g_I Y_a^I- g_{II}Y_a^{II})\sqrt{\frac{2}{N}} \lambda_a 
- i\frac{\delta_{\Sigma}}{2}(g_I Y_a^I\lambda_a\lambda_{\Sigma}-(g_{II}Y_a^{II} \lambda_{\Sigma}\lambda_a)).
\end{eqnarray}
For $\lambda_f$
\begin{eqnarray}
D_{\mu}<\Delta>&=-i\frac{\delta_0}{2}( g_I Y_f^I-g_{II}Y_f^{II})\sqrt{\frac{2}{N}}\lambda_f -i \frac{\delta_{\Sigma}}{2}( g_I Y_f^I-g_{II}Y_f^{II})\nu \sqrt{\frac{2}{N}} \lambda_f \nonumber \\
&=-i \frac{1}{2} \sqrt{\frac{2}{N}}( g_I Y_f^I-g_{II}Y_f^{II})(\delta_0+\nu 
\delta_{\Sigma})\lambda_f.
\end{eqnarray}
From this it follows for the trace $Tr\{(D_{\mu}<\Delta>)^{\dagger} D_{\mu}<\Delta>\}=2\frac{1}{4}\frac{2}{N} (\delta_0+\nu \delta_{\Sigma})^2(g_I^2+g_{II}^2) \bar{Y_f}^2$, $\bar{Y_f}=\frac{1}{\sqrt{g_I^2+g_{II}^2}} (g_IY_f^I-g_{II}^{II})$.\\
For $\lambda_F$
\begin{eqnarray}
D_{\mu}<\Delta>=&-i\frac{\delta_0}{2}( g_I Y_F^I-g_{II}Y_F^{II})\sqrt{\frac{2}{N}}\lambda_F -i \frac{\delta_{\Sigma}}{2}( g_I Y_F^I-g_{II}Y_F^{II})\nu \sqrt{\frac{2}{N}} \lambda_F  \nonumber \\
=& -i \frac{1}{2} \sqrt{\frac{2}{N}}( g_I Y_f^I-g_{II}Y_f^{II})(\delta_0+\nu 
\delta_{\Sigma})\lambda_f,
\end{eqnarray}
similarly $Tr\{(D_{\mu}<\Delta>)^{\dagger} D_{\mu}<\Delta>\}=2\frac{1}{4}\frac{2}{N} (\delta_0+\nu \delta_{\Sigma})^2(g_I^2+g_{II}^2) \bar{Y_F}^2$, \\$\bar{Y_F}=\frac{1}{\sqrt{g_I^2+g_{II}^2}}(g_IY_F ^I-g_{II}^{II})$.\\
For $\lambda_{\Sigma}$
\begin{eqnarray}
D_{\mu}<\Delta>&=-i\frac{\delta_0}{2}( g_I Y_{\Sigma}^I-g_{II}Y_{\Sigma}^{II})\sqrt{\frac{2}{N}}\lambda_{\Sigma} -i \frac{\delta_{\Sigma}}{2}( g_I Y_{\Sigma}^I-g_{II}Y_{\Sigma}^{II})\frac{2}{N}\left( \begin{array}{cc}
\nu^2 & 0 \\
0 & \nu^{-2}\\
\end{array} \right) \qquad \nonumber \\
&= -i \frac{1}{N}( g_I Y_{\Sigma}^I-g_{II}Y_{\Sigma}^{II})
[ \left(\begin{array}{cc}
\delta_0~ \nu & 0 \\
0 & -\delta_0 ~\nu^{-1} \\
\end{array} \right) +
\left( \begin{array}{cc}
\delta_{\Sigma}~\nu^2 & 0 \\
0 & \delta_{\Sigma}~\nu^{-2} \\
\end{array} \right) ]\\
&=-i\frac{1}{N}( g_I Y_{\Sigma}^I-g_{II}Y_{\Sigma}^{II})
[\left( \begin{array}{cc}
\delta_0 ~\nu+\delta_{\Sigma}~\nu^2 & 0 \\
0 & \delta_{\Sigma}~\nu^{-2}-\delta_0 ~\nu^{-1} \nonumber
\end{array} \right) ].
\end{eqnarray}
For the trace
\begin{eqnarray}
Tr\{(D_{\mu}<\Delta>)^{\dagger} D_{\mu}<\Delta>\}=&\frac{1}{N^2}( g_I Y_{\Sigma}^I-g_{II}Y_{\Sigma}^{II})^2 [3n_f (\delta_0 \nu+\delta_{\Sigma}\nu^2)^2+
 \nonumber \\  
N^{'}_cN^{'}_F (\delta_{\Sigma}\nu^{-2}-\delta_0 \nu^{-1})^2]\\
=& N^{'}_cN^{'}_F )(\delta_0 \nu+\delta_{\Sigma}\nu^2)^2 +3n_f(\delta_{\Sigma}\nu^{-2}-\delta_0 \nu^{-1})^2\nonumber \\
=&N \delta_{\Sigma}^2+(\delta_0+(\nu-\bar{\nu}^{'})\delta_{\Sigma})^2,\nonumber
\end{eqnarray}
where was used 
\begin{eqnarray}
\nu^2 &=&\frac{N^{'}_cN^{'}_F}{3N_f},\\
N^{'}_cN^{'}_F \nu^{-1}&=&N^{'}_cN^{'}_F\sqrt{\frac{3n_f}{N^{'}_cN^{'}_F}}=\sqrt{3N^{'}_cN^{'}_F},\\
3n_f \nu&=&3n_f \frac{N^{'}_cN^{'}_F}{3N_f}=\sqrt{3N_f N^{'}_cN^{'}_F },
\end{eqnarray}
for the trace
\begin{equation}Tr\{(D_{\mu}<\Delta>)^{\dagger} D_{\mu}<\Delta>\}=\frac{1}{N}[\delta_{\Sigma}^2+(\delta_0+(\nu-\bar{\nu}^{'}\delta_{\Sigma})^2](g_I^2+g_{II}^2)\bar{Y_\Sigma}^2,
\end{equation}
now for $\lambda_r$
\begin{eqnarray*}
\lambda_r=   
\left( \begin{array}{cc}
0 & 1 \\
1 & 0 \\
\end{array} \right), \qquad \lambda_{r+1}=  
\left[ \begin{array}{cc}
0 & -i \\
i & 0 \\
\end{array} \right],
\end{eqnarray*}
where
\begin{eqnarray*}
\lambda_r\lambda_{\Sigma}=\sqrt{\frac{2}{N}}
\left[ \begin{array}{cc}
0 & 1 \\
1 & 0 \\
\end{array} \right] 
\left[ \begin{array}{cc}
\nu & 0 \\
0 & -\nu^{-1} \\
\end{array} \right]
=\sqrt{\frac{2}{N}} \left[ \begin{array}{cc}
0 & -\nu^{-1} \\
\nu & 0 \\
\end{array} \right] 
\end{eqnarray*}

\begin{eqnarray*}
\lambda_{\Sigma}\lambda_r=\sqrt{\frac{2}{N}}
\left[ \begin{array}{cc}
\nu & 0 \\
0 & -\nu^{-1} \\
\end{array} \right] 
\left[ \begin{array}{cc}
0 & 1 \\
1 & 0 \\
\end{array} \right] 
=\sqrt{\frac{2}{N}} \left[ \begin{array}{cc}
0 &\nu  \\
-\nu^{-1} & 0 \\
\end{array} \right] 
\end{eqnarray*}

\begin{eqnarray*}
\lambda_{r+1}\lambda_{\Sigma}=\sqrt{\frac{2}{N}}
\left[ \begin{array}{cc}
0& -i \\
i & 0  \\
\end{array} \right]
\left[ \begin{array}{cc}
\nu & 0 \\
0 & -\nu^{-1} \\
\end{array} \right] 
=\sqrt{\frac{2}{N}} \left[ \begin{array}{cc}
0 & -i \nu^{-1}  \\
i \nu & 0 \\
\end{array} \right]
\end{eqnarray*}

\begin{eqnarray*}
\lambda_{\Sigma}\lambda_{r+1}=\sqrt{\frac{2}{N}}
\left[ \begin{array}{cc}
0& -i\nu \\
-i \nu^{-1} & 0 \\
\end{array} \right],
\end{eqnarray*}
for the trace 
\begin{eqnarray}& &
Tr\{(D_{\mu}<\Delta>)^{\dagger} D_{\mu}<\Delta>\}=\frac{1}{N}(\delta_0 +\nu \delta_{\Sigma})^2 Y_f^2+ 
\frac{1}{N}(\delta_0 -\nu^{-1} \delta_{\Sigma})^2 (g_I^2+g_{II}^2)\bar{Y_F}^2 
+ \nonumber \\ & &
\frac{1}{N}[\delta_{\Sigma}^2 +(\delta_0 +(\nu-\bar{\nu}^{'})\delta_{\Sigma})^2](g_I^2+g_{II}^2)\bar{Y_{\Sigma}}^2 
+ \frac{1}{N}[\delta_0 +\frac{1}{2}(\nu-\bar{\nu}^{'})\delta_{\Sigma})^2 \Sigma_{\epsilon}(g_I Y_{\epsilon}^I-g_{II}Y_{\epsilon})^2 
+ \nonumber \\ & &
\frac{1}{4N}\delta_{\Sigma}^2 (\nu-\bar{\nu}^{'})^2 \sum_{\epsilon}(g_I Y_{\epsilon}^I+g_{II}Y_{\epsilon}^{II})^2.
\end{eqnarray}
To obtain the mass matrix of bosons $Y_i$ is now necessary to consider all possible paths of spontaneous symmetry breaking 
$G_H\to SU(3)_c\otimes SU(N^{'})_{c^{'}}$
through $<\Delta>\neq$0 is broken to
\begin{equation}
SU(N)_I\otimes SU(N)_{II} \to SU(N)_{I+II},
\end{equation}
\begin{equation}
SU(N)_{L+dR+uR}\to SU(3)_c\otimes SU(N^{'}_{c^{'}}),
\end{equation}
now we have
\[a(g_IY^I-g_{II}Y^{II})^2+b(g_IY^I+g_{II}Y^{II})^2-2ag_Ig_{II}Y^IY^{II}+2b g_Ig_{II}Y^IY^{II} (2(b-a))\]
\[
\left( \begin{array}{cccccc}
a_{11} &0& a_{13} &0 & a_{15} &0\\
0 & a_{22} &0& a_{24} &0 & a_{26}\\
a_{13} &0& a_{33} &0& a_{35} &0\\
0 & a_{24} &0& a_{44} &0 & a_{46}\\
a_{15} &0& a_{35} &0& a_{55} &0\\
0 & a_{26} &0& a_{46} &0 & a_{66}\\
\end{array} \right). 
\]
From the eigenvalue problem
\[
\left[ \begin{array}{cccccc}
a-\lambda&0& b,&0& c&0\\
0 &d-\lambda&0&e&0 &f\\
b&0& g-\lambda&0& h&0\\
0 &e&0&i-\lambda&0 &j\\
c&0& h,&0& k-\lambda&0\\
0 &f&0&j&0 &l-\lambda\\
\end{array} \right], 
\]
solving
\[
(a-\lambda)
\left[ \begin{array}{ccccc}
d-\lambda & 0& e&0& f\\
0 & g-\lambda &0&h&0\\
e&0& i-\lambda &0& j\\
0 &h&0& k-\lambda &0 \\
f&0& j&0& l-\lambda\\
\end{array} \right] 
+
b
\left[ \begin{array}{ccccc}
0& d-\lambda& e&0& f\\
b &0 &0&h&0\\
0& e& i-\lambda &0& j\\
c &0&0& k-\lambda &0 \\
0&f& j&0& l-\lambda \\
\end{array} \right] 
\]

\[
+c
\left[ \begin{array}{ccccc}
0& d-\lambda & 0&e& f\\
b &0 & g-\lambda &0&0\\
0& e&0& i-\lambda & j\\
c &0&h&0&0 \\
0&f&0& j& l-\lambda\\
\end{array} \right]. 
\]
We have $Y_i$ $i=1,2,...6$; $Y_i=a_{il}\bar{Y_l}$, is found that $Y_{odd}$ does not mix with $Y_{even}$ thus the breaking $G_H\to G_{Min}$ can not mix them. For the other breaks
\begin{equation}
SU(N)_i\to SU(3)_{ic}^{f_1}\otimes SU(3)_ {ic}^{f_2}\otimes SU(N^{'})_ {ic^{'}}^{f_{\phi}},
\end{equation}
through the higgses $\chi_{[123]}$, $\chi_{[456]}$ being found not mix between the $Y_i$ thus when used $\chi_{[ijk]}$ here is no additional mixing of the $Y_i^{ETC}$.
\begin{equation}
SU(N)_I\otimes SU(N)_{II} \to SU(N)_{I+II},
\end{equation}
realized through the $<\Delta>=\delta_0 \lambda_0$.
Here there exist a mix in the form 
\begin{equation}
 \frac{1}{N}\delta_0^2(g_IY_I-g_{II}Y_{II})^2=c (g_I^2Y_I^2-2g_Ig_{II}Y_IY_{II}+g_{II}^2 Y_{II}^2).\end{equation}
Besides
\begin{equation}
SU(N)_I\otimes SU(N)_{II}\to SU(3n_f)_{I+II}\otimes SU(N^{'}_cN^{'}_F)_{I+II}\otimes U(1)_{\Sigma}^{I+II},
\end{equation}
through the election $<\Delta>=\delta_0\lambda_0 +\delta_{\Sigma}\lambda_{\Sigma}$; mixture exists $a(g_IY_I-g_{II}Y_{II})^2 +b(g_IY_I+g_{II}Y_{II})^2$, $a=\frac{1}{N}(\delta_0+1/2(\nu-\nu^{'}\delta_{\Sigma})^2$, $b=\frac{1}{4N}(\nu+\bar{\nu^{'}})^2 \delta_{\Sigma})^2$, where $a=b$ if $\lambda_0\sim\lambda_{\Sigma}\neq 0$. The mixture is greater in case b.
The situation with $N_f$ families is such that it has
\[
 \begin{array}{cccc}
Y_1 & Y_{n_{f+1}} &  Y_{2n_{f+1}} & \to mixing\\
\cdot&\cdot& \cdot& \to mixing\\
\cdot&\cdot& \cdot& \to mixing\\
\cdot&\cdot& \cdot& \to mixing\\
Y_f & Y_{2n_f} &  Y_{3n_f} & \to mixing, \\ 
\end{array} 
\]
and there will be to solve a cubic equation for the eigenvalues. We then obtain after much calculate the branching of the form\\
$G_H=SU(N)_L^q\otimes SU(N)_R^u\otimes SU(N)_R^d\otimes SU(M)_L^l \otimes SU(M)_R^e \times G_V$
\begin{eqnarray}
G_1&=&G_H \nonumber \\
G_2&=&SU(N)_{q_L+u_R}\otimes \dots \nonumber \\
G_3&=&SU(N)_{q_L+d_R}\otimes \dots\\
G_4&=&SU(N)_{q_L}\otimes SU(N)_{u_R+d_R}\otimes \dots \nonumber \\
G_5&=&SU(N)_{q_L+u_r+d_R}, \nonumber
\end{eqnarray}
con $G_{UB}=SU(3)_c\otimes SU(N^{'}_c)_{c^{'}}$ (see Figure \ref{fig}).
\section{Discussion}
It remains to continue the implementation for the biological case, where according to \cite{bashfords1} could introduce a symmetry similar to isospin in nuclear physics and that has given the name of polarity. This reinforces our intention to use the tools of particle physics in the algebraic approximation of Lie groups introduced by Hornos \cite{main,PRL71,PRE} to study the structure of the genetic code and in particular certain specific symmetries as $sp(6)$ to reproduce the pattern of degeneration of the codons in multiplets. Moreover the fact that Forger has involved already Lie superalgebras 
\cite{PRE}, and Brashford \cite{bashfords,bashfords1} too, and the fact that due to degeneration when making  assignment of codons to a few amino acids, which requires enlarge more the symmetry, when this symmetry is broken to subalgebras, this gives rise to the need to incorporate more amino acids each with their own redundancy in the codons. In \cite{bashfords3} was emphasized that biochemical factors were key elements for studying the evolution of the genetic code. Here is that the incorporation of the horizontal symmetry could give us information regard it, because it could study the labeling scheme. Then has to be find the pattern which leads us to find a correct content of multipletes corresponding to the genetic code previously calculated with the sp(6) symmetry. Like in the great unification models we would have here a pattern of repetition of families.  

\end{document}